\documentclass[12pt,preprint]{aastex}
\usepackage{amssymb,amsmath,amstext,amsgen,amsopn,amsxtra,indentfirst}
\shorttitle{Dust Echoes from GRBs}
\shortauthors{Heng, Lazzati \& Perna}
\begin{document}

\title{DUST ECHOES FROM THE AMBIENT MEDIUM OF GAMMA-RAY BURSTS}

\author{Kevin Heng\altaffilmark{1}, Davide Lazzati\altaffilmark{1} 
\& Rosalba Perna\altaffilmark{1}}

\altaffiltext{1}{JILA, University of Colorado, 440 UCB, Boulder, 
CO 80309-0440; hengk@colorado.edu}

\begin{abstract}
Long gamma-ray bursts (GRBs) are likely associated with the collapse
of massive stars, which produce dust and are born in dusty
environments.  Absorption and scattering of ultraviolet/X-ray photons
from the prompt, optical flash and afterglow emission of the GRB
produce dust echoes.  We perform time-dependent calculations of these
echoes, accounting for the evolution of the dust grain distribution
due to selective grain destruction by the GRB radiation, and for
off-axis beaming.  We explore cloud configurations of differing density and size --- the echo light curve and
spectrum depend on the cloud radius, with larger clouds peaking at
longer wavelengths.  For a region $\sim3$ pc in size with density
$n_{\rm{H}} \sim 10^3$ cm$^{-3}$, the echo spectrum peaks at $\sim$ 3.6 $\mu$m and $\sim 8.8$ eV for thermal and scattered components, respectively.  Dust echoes should be detectable with the {\it Very Large Telescope} up to $z \sim 0.1$, IRAC onboard the {\it Spitzer Space Telescope} up to $z \sim 0.2$, and NICMOS onboard the {\it Hubble Space Telescope} up to $z \sim 0.3$.  Furthermore, the shape of the echo light curve allows one to infer: the jet opening angle; the inclination of the jet axis with respect to the line of sight; the size of the dust-emitting region.  For sources with
symmetric, bipolar jets, dust echoes exhibit two bumps in the light
curve, making them easily distinguishable from the rebrightening due
to an underlying supernova.
\end{abstract}


\section{INTRODUCTION}
\label{sect:intro}
Dust echoes are produced when photons emitted from a source (nova,
supernova, quasar, gamma-ray burst, etc) interact with the ambient
dust grains, which either scatter or thermally re-emit the photons and
divert some of them towards the observer.  These photons may be
observed long after the source has faded, thus providing a way of
studying both the nature of the source and its surroundings.  A number
of papers have been written on dust echoes, including those by Dwek
(1983), Chevalier (1986), Milgrom (1987), Schaefer (1987a, b),
Emmering \& Chevalier (1989), Sparks (1994), Xu, Crotts \& Kunkel
(1994, 1995), Sugerman (2003) and Patat (2005, hereafter P05),
following pioneering work by Couderc (1939). All of these papers deal
with the dust echoes --- also called ``light echoes'' --- produced by
static dust distributions illuminated by a relatively weak flux of
radiation.  In gamma-ray bursts (GRBs), the radiation flux is so large
that dust can be destroyed out to $\sim$ 10 pc by ultraviolet (UV) and
X-ray heating, ion field emission (IFE; Waxman \& Draine 2000,
hereafter WD00) or Coulomb explosions (Fruchter, Krolik \& Rhoads
2001).

The modeling of dust echoes in GRBs is therefore complicated by the
need to perform a time-dependent treatment of the dust, where
temperatures and size distributions evolve as a function of time and
distance from the source.  Dust grains are selectively destroyed
according to their size and chemical composition, making it impossible
to define a destruction radius uniquely.  Previous work either relied
on neglecting the dust evolution or assumed a destruction radius
within which all of the dust is destroyed and beyond which it is
unaffected.  These include the work of WD00, Esin \& Blandford (2000,
hereafter EB00), Reichart (2001, hereafter R01), Venemans \& Blain
(2001) and Moran \& Reichart (2005, hereafter MR05).

The motivation for considering GRB dust echoes originates from
researchers wanting to provide an alternative interpretation of the
evidence for a GRB/supernova connection (e.g., WD00; EB00; R01) and to
explain color fluctuations in afterglow light curves (MR05).  With the
discovery of the association between long GRBs and the collapse of
massive stars (Hjorth et al. 2003; Stanek et al. 2003), the
alternative explanation for supernova (SN) bumps is no longer
necessary.  However, since massive stars have relatively short
lifetimes, the majority of them are expected to die within the dense
and dusty molecular clouds in which they were created.  Dense
environments, with conditions comparable to those in molecular cloud
cores, have indeed been inferred in a few cases (Lazzati \& Perna
2002; Frontera et al. 2004; Frail et al. 2006).  The study of dust
echoes from GRBs then becomes motivated by the fact that they provide
a three-dimensional map of the geometry of the emitting region.  For an
isotropic dust distribution, this in turn translates into a geometric
mapping of the source of radiation, constituting an independent way of
constraining the degree of beaming of GRB jets.

In this paper, we present significant improvements to existing models
of dust echoes by combining time-dependent echo calculations, which
are able to account for off-axis, bipolar GRB beaming, with a code
that follows the evolution of the dust grain population.  The
photoionization and dust destruction code, written by Perna \& Lazzati
(2002, hereafter PL02; see also Perna, Lazzati \& Fiore 2003 and
Lazzati \& Perna 2003), models the destruction of dust grains due to
UV and X-ray sublimation, as well as IFE.  Evolution of the (graphite
and silicate) grains, hydrogen and the 12 most abundant metals are
self-consistently combined.  The mono-dimensional output of the dust
code is subsequently fed to a three-dimensional echo code that
computes the contribution of the dust populations to the thermal and
scattered echoes.  In \S\ref{sect:methods}, we discuss the dust echo
geometry and GRB parameters adopted, and describe our use of the PL02
code.  We present our results in \S\ref{sect:results} and discuss
their implications in \S\ref{sect:discussion}.

\section{METHODS}
\label{sect:methods}

\subsection{ECHO GEOMETRY}
\label{subsect:lightecho}

\begin{figure}
\begin{center}
\includegraphics[width=6in]{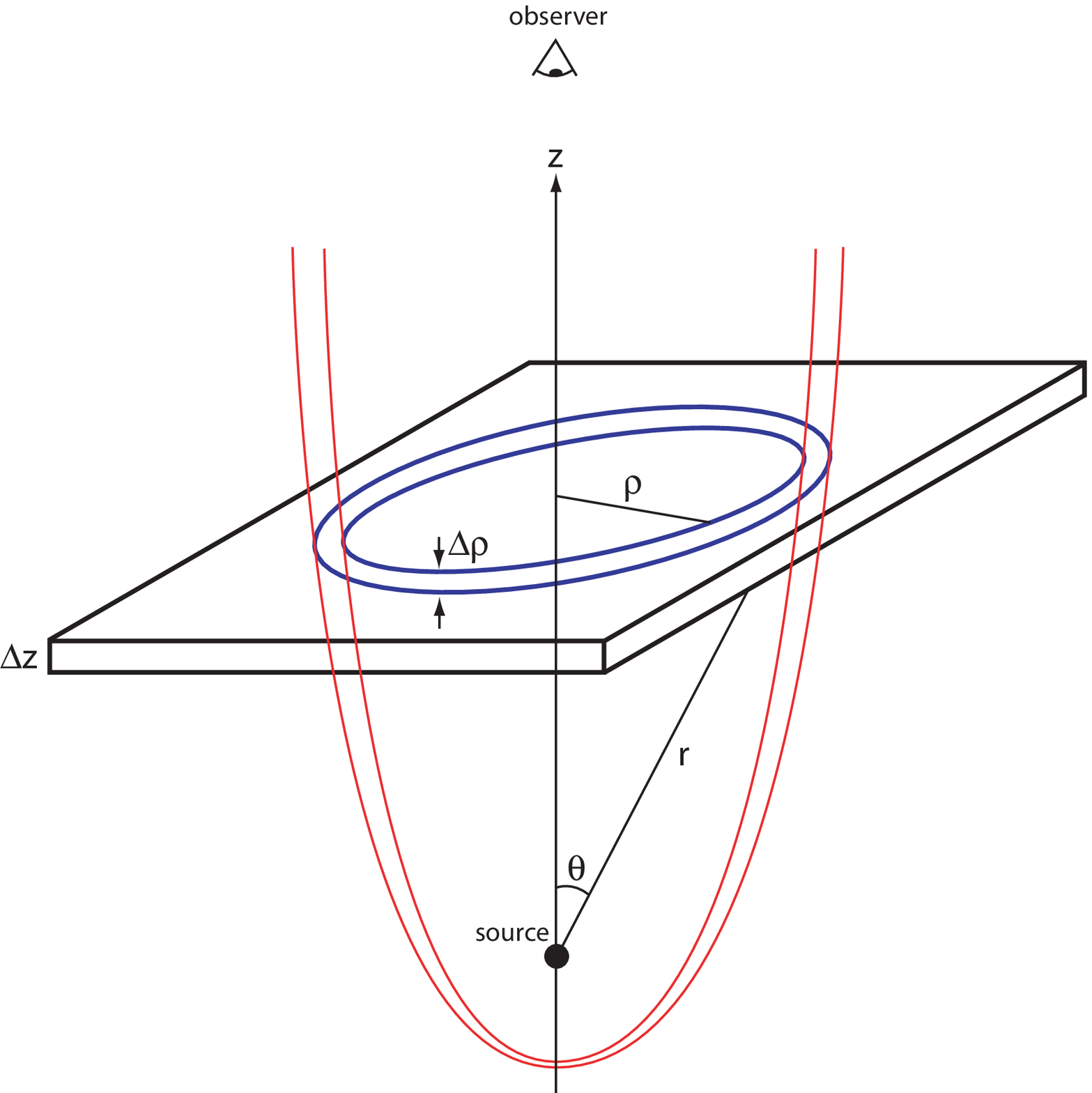}
\end{center}
\caption{Dust echo geometry, adopted from Sugerman (2003).}
\label{fig:geometry}
\end{figure}

We adopt the dust echo geometry of Sugerman (2003) and consider a
coordinate system in which a thin, planar sheet of dust, of thickness
$dz$, is located a distance $z$ away from the source
(Fig. \ref{fig:geometry}).  From the point of view of the observer, an
echo of projected radius $\rho$ and width $d\rho$ is seen.  Dust
grains, each located a distance $r$ from the source, scatter the light
onto parabolic surfaces (Couderc 1939):
\begin{equation}
z = \frac{\rho^2}{2 c t} - \frac{c t}{2}.
\end{equation}

Suppose an isotropic burst of energy (per unit frequency),
$E_{\nu,\rm{input}} = E_{\nu,\rm{input}}(\nu,t)$, is released by the
source at a time $t_0$, over some (global) time interval $\delta t = t
- t_0$ and at a frequency $\nu$, such that $L_{\nu,\rm{input}} =
E_{\nu,\rm{input}}/\delta t$.  We first calculate the integrated,
monochromatic luminosity, $L_{\nu,\rm{sc}}=L_{\nu,\rm{sc}}(\nu,t)$, due to
scattering, at some time $t$ after the burst.  If $a$ denotes the
grain size and $n = n(a,r,t)$ is the dust density,
\begin{equation}
L_{\nu,\rm{sc}} = \int \int_V \frac{L_{\nu,\rm{input}} ~a^2 ~Q_{\rm{sc}} ~\Phi}{4
~r^2} ~\frac{dn}{da} ~\frac{\delta t}{\Delta t} ~dV ~da,
\end{equation}
where $Q_{\rm{sc}}=Q_{\rm{sc}}(a,\nu)$ is the scattering efficiency and $dV = 2
\pi \rho ~d\rho ~dz$ is the echo volume element in the absence of
beaming. The {\it local} time interval, $\Delta t$, is the time lag
between successive echo parabolic surfaces.  In practice, we follow
the evolution of the dust grains in discrete bins of $\Delta a$ (see
\S\ref{subsect:dustcode}).  The Henyey \& Greenstein (1941, hereafter
HG41) phase function takes the form
\begin{equation}
\Phi = \frac{1 - g^2}{(1 + g^2 - 2g\mu)^{3/2}},
\end{equation}
with $g = g(a,\nu)$ characterizing the degree of forward scattering
for a given grain.  The quantity $\mu = z/r$ is the cosine of the
scattering angle, $\theta$.  In almost all of the work cited, the HG41
phase function is adopted.  P05 notes that more sophisticated
calculations, such as a full Mie treatment, show that the HG41
formulation tends to underestimate the forward scattering, but the
approximation is reasonably good.

For thermal re-emission, the integrated, monochromatic luminosity is
\begin{equation}
L_{\nu,\rm{th}} = \int \int_V 4 \pi^2 ~a^2 ~B_\nu ~Q_{\rm{abs}} ~\frac{dn}{da}
~\frac{\delta t}{\Delta t} ~dV ~da,
\end{equation}
where $Q_{\rm{abs}} = Q_{\rm{abs}}(a,\nu)$ is the absorption efficiency and
$B_\nu = B_\nu(T(a,r,t),\nu)$ is the Planck or blackbody intensity.
Collectively, the monochromatic light curve observed is given by
\begin{equation}
L_\nu = L_{\nu,\rm{sc}} + L_{\nu,\rm{th}}.
\end{equation}

Though we use a planar slab for the purpose of illustration, our
formalism can be applied to any geometry.  Analytical formulae for the
simple cases of a slab and a spherical shell are given in, for
example, P05.  Tabulated values for $g$, $Q_{\rm{sc}}$ and $Q_{\rm{abs}}$ are
provided by Draine \& Lee (1984) and Laor \& Draine (1993).

\begin{figure}
\begin{center}
\includegraphics[width=6in]{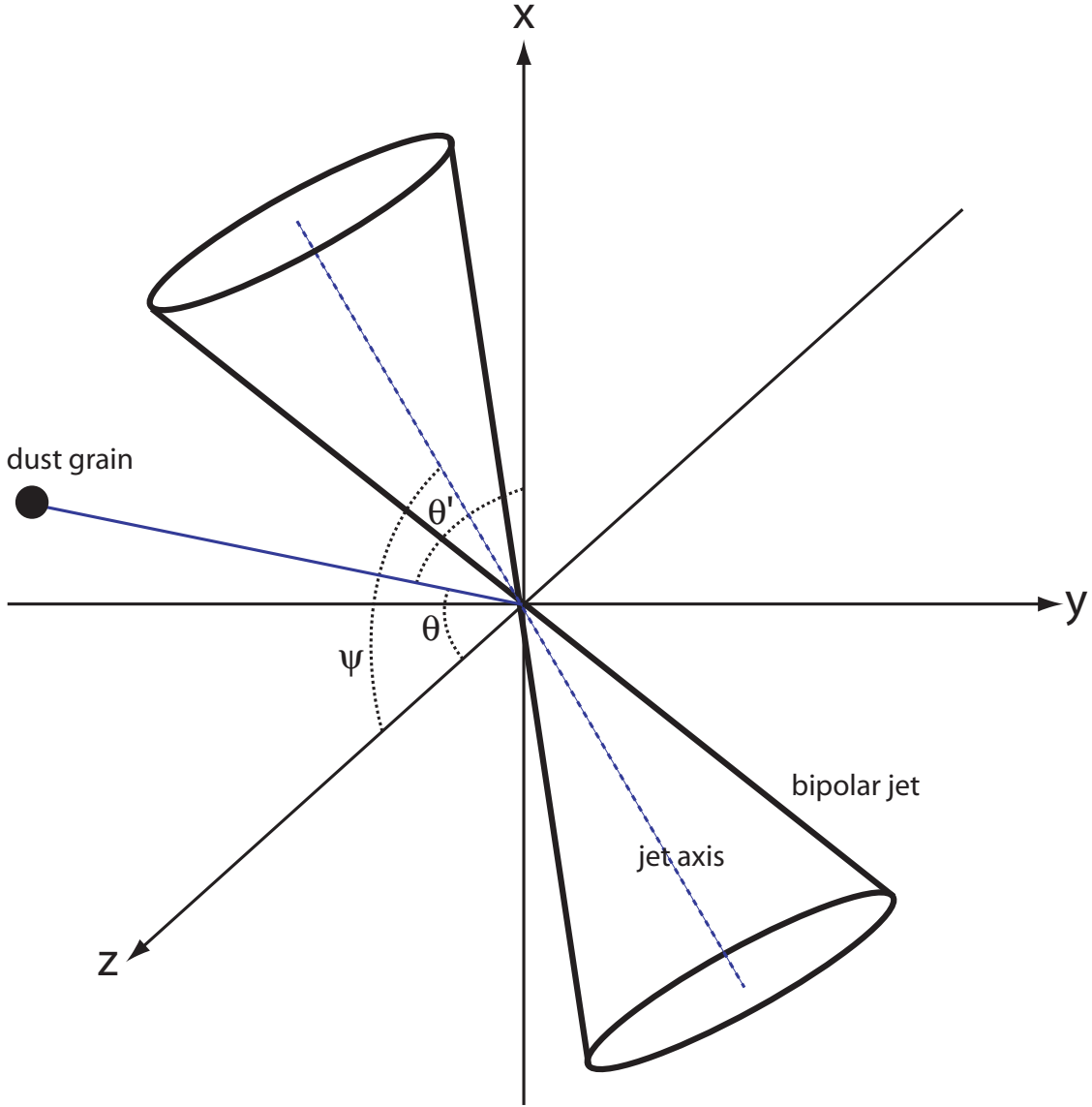}
\end{center}
\caption{Geometry for GRB beaming.  A bipolar jet of opening half-angle $\theta_{\rm{jet}}$ (not displayed in the diagram) is positioned at a viewing angle of $\psi$ between the jet axis and the light of sight ($z$-axis).  While the jet axis is located in the $x$-$z$ plane, the dust grain shown is generally not found in the same plane; $\theta$ is the scattering angle.  Knowledge of the occurrence angle, $\theta^\prime$, allows one to determine if the light curve contribution from a given dust grain should be included (see text).  Also not shown is the angle $\phi$, which is the azimuthal angle about the line of sight.}
\label{fig:beaming}
\end{figure}

In our case, we consider a bipolar jet of opening half-angle
$\theta_{\rm{jet}}$, positioned at a viewing angle $\psi$ between the
jet axis and the line of sight (Fig. \ref{fig:beaming}).  The cosine
of the occurrence angle, $\theta^\prime$, is given by
\begin{equation}
\cos{\theta^\prime} = \cos{\theta} \cos{\psi} + \sin{\theta} \sin{\psi} \cos{\phi}.
\end{equation}
The luminosity contribution from a given dust grain is only included
when either the condition $\theta^\prime < \theta_{\rm{jet}}$ or
$\theta^\prime > \pi - \theta_{\rm{jet}}$ is satisfied.  The echo
volume element is then generalized to $dV = \rho ~d\phi ~d\rho ~dz$
and there is an additional integration over the azimuthal angle,
$\phi$.  Again, in practice, one sums over discrete intervals or bins
of $\Delta \rho$, $\Delta \phi$, $\Delta z$ and $\Delta a$.

\subsection{GRB LIGHT CURVE MODELING}
\label{subsect:grbparameters}

\begin{figure}
\begin{center}
\includegraphics[width=6in]{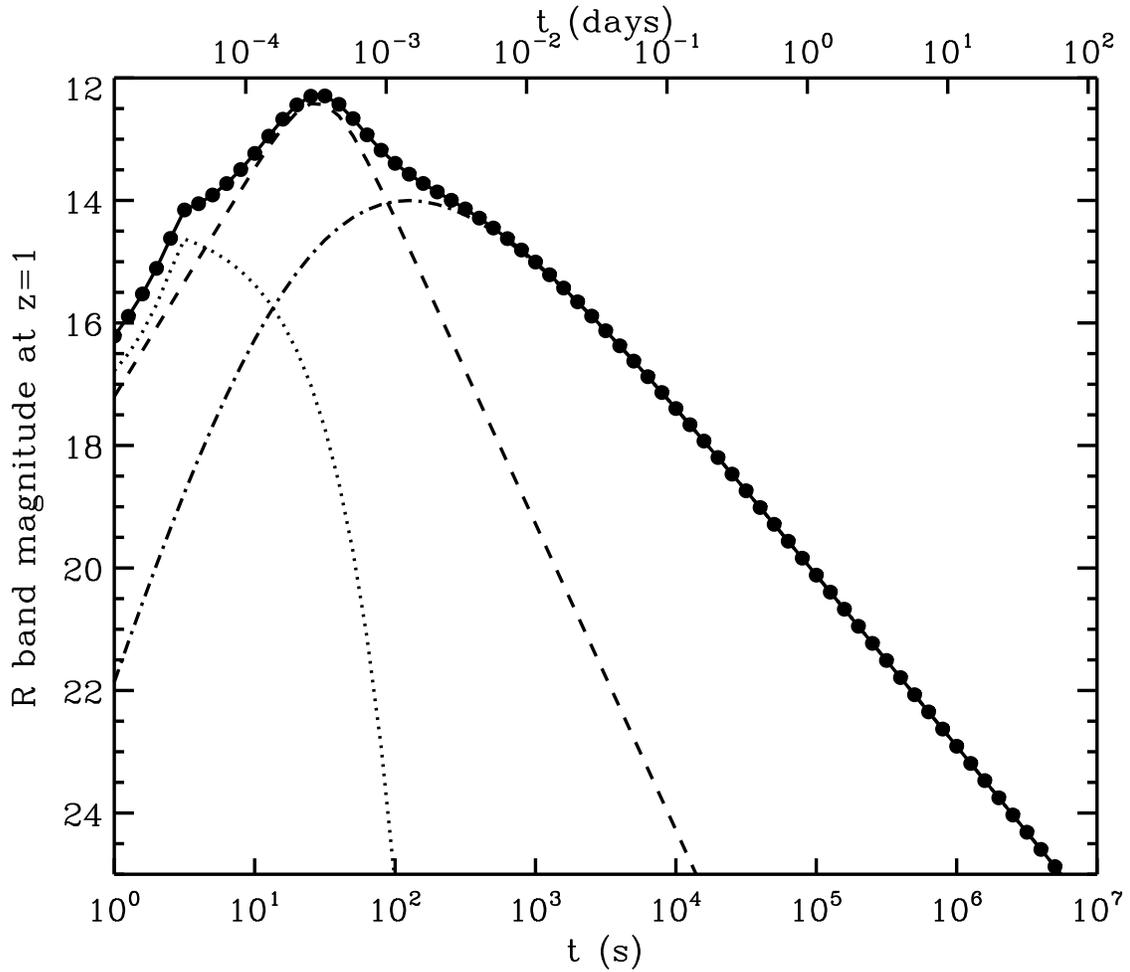}
\end{center}
\caption{{\it R} band light curve according to our GRB model for an event at
$z=1$. The solid line with dots shows the total light curve, while
dotted, dashed and dot-dashed lines represent the prompt, optical flash, and
afterglow emission, respectively.
\label{fig:lc}}
\end{figure}

The GRB light curve has three components: the prompt emission, an
optical flash and the afterglow (Fig. \ref{fig:lc}).  We model the prompt emission as
\begin{equation}
L_{\nu,\rm{pm}}(\nu,t) = \frac{L_0}{1+\left(\nu/6 \times 10^{19} \mbox{ Hz}\right)^{2.5}}
\begin{cases}
\exp{\left({-3 + t/1 \mbox{ s}}\right)} &, t \leq 3 \mbox{ s}\\
\exp{\left({0.3 - t/10 \mbox{ s}}\right)} &, t \geq 3 \mbox{ s}.
\end{cases}
\end{equation}
The quantity $L_0 = 1.16 \times 10^{32}$ erg s$^{-1}$ Hz$^{-1}$ is
chosen such that $\int \int L_{\nu,\rm{pm}} ~d\nu ~dt \approx 10^{53}$ erg.

We allow for the presence of an optical flash, as observed in GRB
990123 (Akerlof et al. 1999), modeled through the function (e.g.,
Draine \& Hao 2002)
\begin{equation}
L_{\nu,\rm{of}}=3\times10^{33} \mbox{ erg s}^{-1} \mbox{ Hz}^{-1} ~\frac{(t/30 \mbox{ s})^{3/2}}{1 + (t/30 \mbox{ s})^{7/2}}
\begin{cases}
\left(\nu/4 \times 10^{14}\mbox{ Hz}\right)^{-0.65} &, \nu < \nu_B\\
100^{-0.65} ~\exp{\left(\nu/\nu_B-1\right)} &, \nu \geq \nu_B
\end{cases}
\end{equation}
where $\nu_B=8\times10^{17}$~Hz.

For the afterglow, we use (Panaitescu \& Kumar 2000)
\begin{equation}
L_{\nu,\rm{ag}} = 3 \times 10^{33} \mbox{ erg s}^{-1} \mbox{ Hz}^{-1}
\left[\left(t/30 \mbox{ s}\right)^{-0.5} +
\left(t/30 \mbox{ s}\right)^{1.125/6}\right]^{-6} ~\left( \frac{\nu}{4
\times 10^{14} \mbox{ Hz}} \right)^{-1.15}.
\end{equation}

It follows that the input luminosity (per unit frequency) from the GRB is
\begin{equation}
L_{\nu,\rm{input}} = L_{\nu,\rm{pm}} + L_{\nu,\rm{of}} +L_{\nu,\rm{ag}}.
\end{equation}
Figure~\ref{fig:lc} shows the {\it R} band light curve according to our
model for a GRB at $z=1$. If the GRB ejecta are beamed, then the
afterglow eventually evolves to $L_{\nu,\rm{ag}} \propto t^{-p}$, where $p
= 2.5$ is the index of the electron energy distribution, $F(\epsilon)
\propto \epsilon^{-p}$, at a time
\begin{equation}
t_{\rm{break}}(\theta_{\rm{jet}}) = \left( \frac{3E}{4 \pi c^5 m_H n}
\right)^{1/3} ~\theta^{8/3}_{\rm{jet}},
\end{equation}
where $E = 10^{53}$ erg and $n$ is the number density of the ambient
gas.  In the case of an off-axis jet, we adopt the following
approximation: at $t_{\rm{break}} = t_{\rm{break}}(\psi)$ (Rossi,
Lazzati \& Rees 2002; Zhang \& M\'{e}sz\'{a}ros 2002), $L_{\nu,\rm{ag}}$
transitions from $\propto t^2$ to $\propto t^{-p}$; we ignore the
$\propto t^3$ and plateau behaviors just before $t_{\rm{break}}$.

\subsection{DUST MODELING}
\label{subsect:dustmodel}

As discussed in \S1, the massive-star progenitors of GRBs --- probably
of the Wolf-Rayet type --- are likely to die close to the locations
where they were born, in dense, molecular cloud cores.  While
we compute the peaks of the dust echo emission for different cloud configurations, we focus most of the calculations
on a typical example of such a cloud; specifically, one with radius
$R=3$ pc and hydrogen column density $N_{\rm{H}}=10^{22}$ cm$^{-2}$,
corresponding to $n_{\rm{H}} \sim 10^{3}$ cm$^{-3}$ or, equivalently, a gas mass
of $\sim$ 1000 $M_\sun$.  The GRB is assumed to be located at the
center of the cloud. Since late-type WC stars are
themselves known to be dust-making machines (e.g., Williams, van der
Hucht \& The 1987), a further concentration of dust is expected in the
vicinity of the star.  Observations by Williams, van der Hucht \& The
(1987) indicate these distances to be $\sim$ 10$^{14}$ to 10$^{15}$
cm.  MR05 considered specifically the emission from this proximate
dust layer, showing that it yields a peak emission within a few
minutes after the burst.  We consider a uniform distribution of dust
within the molecular cloud, with a solar dust-to-gas mass ratio,
$f_\sun = 0.01$.

A variety of grain models exist and the interested reader is referred
to, for example, the review by Draine (2004).  In general, they
incorporate a combination of silicates, graphite and polycyclic
aromatic hydrocarbons (PAHs).  Broadly speaking, the presence of
graphite grains may account for the 2175 \AA~ bump in the extinction
curve.  Steep UV extinction may be due to the presence of silicate
grains.  PAHs account for several unique emission features, as well as
for far ultraviolet extinction. Various types of composite
grains may also provide the bulk of the infrared and optical
extinction.  We specialize to Galactic-type dust, which is believed to
be a mixture of graphite and silicate grains --- with the latter
having an average composition of MgFeSiO$_4$, also known as
``olivine''; see Draine \& Lee (1984) --- following Mathis, Rumpl \&
Nordsieck (1977).  We assume an initial, power-law distribution
\begin{equation}
\frac{dn}{da} \propto a^{-\beta},
\end{equation}
where $\beta = 3.5$.  While this value of $\beta$ is typical of the
dust found in our Galaxy, it might not accurately represent the grain
distributions in denser environments.  Nevertheless, it is a good
first approximation.  As the burst radiation propagates through the
medium, the grain population evolves and $\beta$ either changes or
ceases to be meaningful as portions of the distribution may be
completely destroyed.  Dust echoes probe the distribution of dust
while it is being modified by the burst radiation. We properly compute
the temporal evolution of the dust distribution, as described in the
succeeding sub-section.

\subsection{DUST EVOLUTION}
\label{subsect:dustcode}

As mentioned in \S\ref{sect:intro}, we follow the evolution of dust as
a function of position within the cloud by using the dust code of
PL02, which treats metals (He, C, N, O, Ne, Mg, Si, S, Ar, Ca, Fe,
Ni), hydrogen (in all of its atomic and molecular forms) and dust
evolution in a self-consistent manner.  For dust destruction, UV and
X-ray sublimation and IFE are considered.  Sublimation occurs when
grains absorb energy faster than they manage to radiate away and the
excess energy goes into breaking the bonds that hold the atoms to the
surface of the grain.  Small and large grains are heated predominantly
by UV radiation and X-rays, respectively.

IFE occurs when a grain is charged up to a potential of
$V_{\rm{max}}=3 a_{-5}(S/10^{11}$ dyn cm$^{-2})^{1/2}$ kV, where $a =
a_{-5} 10^{-5}$ cm and $S$ is the tensile strength of the grain
material, and any further ionization leads to the emission of ions in
order to avoid an increase in the surface electric field.  The process
of Coulomb explosion, which is the fragmentation of the grain into smaller
components, competes with that of IFE.  The relative importance of the
two processes is a subject of controversy.  PL02 adopt the approach of
Draine \& Hao (2002) and assume that Coulomb explosion does not play a major role, due to the possibility of chemical bonds being promptly re-established
in the warm grain --- annealing.  Grains close to the source are
sublimated, while those far away from it are eroded by IFE.

\begin{figure}
\begin{center}
\includegraphics[width=6in]{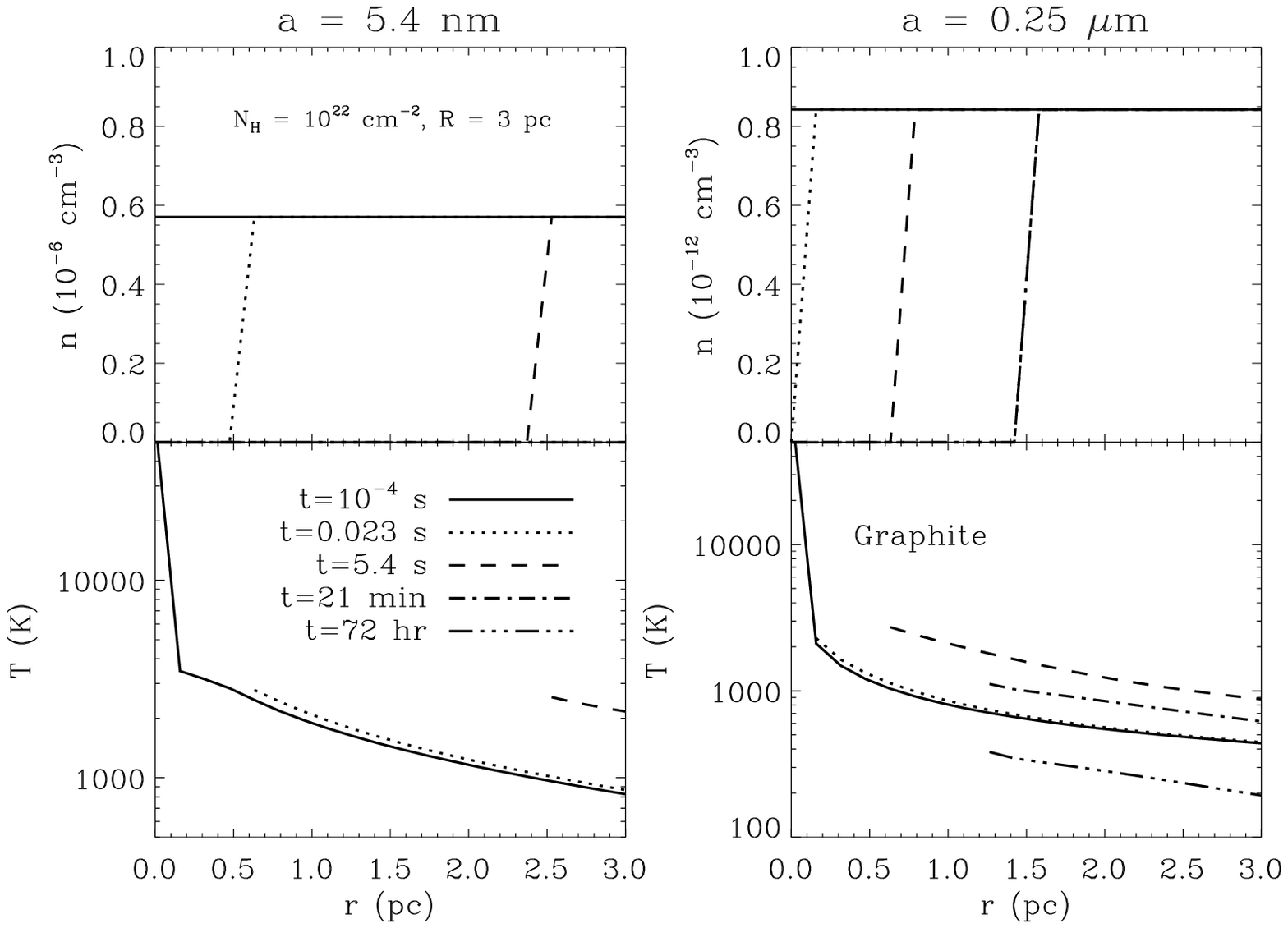}
\includegraphics[width=6in]{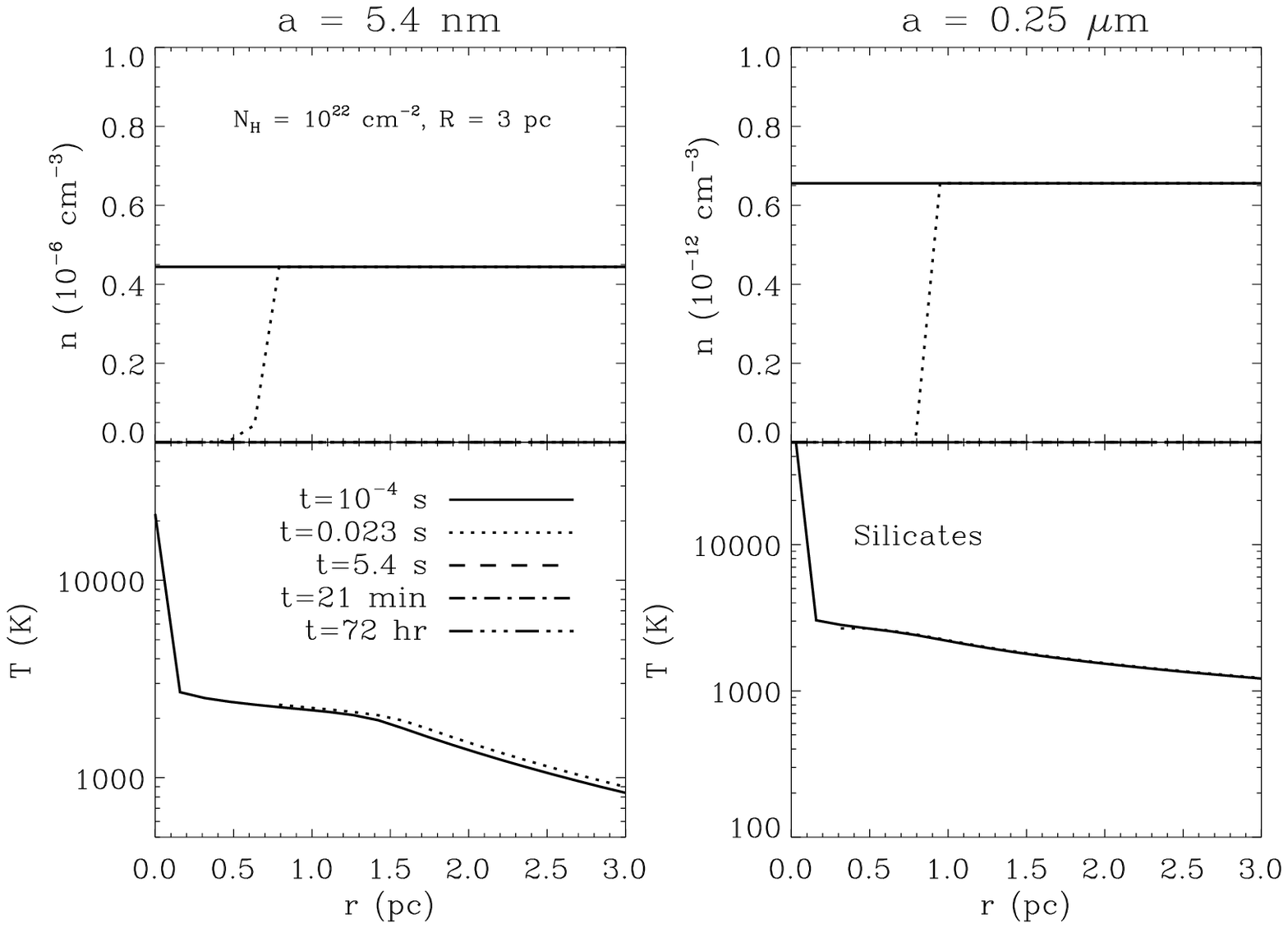}
\end{center}
\caption{Input number densities, $n$, and temperatures, $T$, for two representative grain sizes and at different times since the GRB, generated using the dust code of Perna and Lazzati (2002).  Shown are graphite (top) and silicate (bottom) grains.}
\label{fig:inputs}
\end{figure}

As dust destruction fronts propagate, dust-depleted metals are
recycled into gas.  As such, their abundances in the gaseous and the
dust-depleted phases are book-kept separately.  Computationally, at
each time step $t$, the code calculates $dn/da$ and $T$ as a function
of both $r$ and $a$ [Fig. \ref{fig:inputs}; see also Perna, Lazzati \&
Fiore (2003) for further examples of dust evolution for various
initial values of $\beta$]. In this way, the dust populations are
evolved both spatially and temporally.  We see from
Fig. \ref{fig:inputs} that graphite grains are more resistant to the
impinging radiation; larger grains tend to survive better as well.

\subsection{CLOUD OPACITY}

Neither the dust code of PL02 nor our echo code
deals with the multiple-scattering effects of radiative transfer in an
optically-thick cloud. There are three regimes of opacities, namely clouds that are: optically thin at the beginning of the burst; optically thick at the end of the burst and afterglow; initially optically thick but become optically thin after part of
the dust is destroyed by the burst and afterglow radiation. Certainly, this
definition is dependent on wavelength.

Clouds that are initially optically thin to radiation are not a
concern, since multiple scattering and absorption/re-emission are
negligible. However, a cloud can be optically thin in the infrared but
thick in the UV; in principle, UV radiation can be ``downgraded'' to the infrared
by multiple scattering and absorption/re-emission events, thus contributing to the infrared lightcurve.  However, this effect is small for two
reasons. Firstly, the dust temperatures at long time scales are low, due to the rapid cooling of the grains.  UV radiation will be reprocessed in the far
infrared, at wavelengths $\sim$ 10 $\mu$m, much longer than those considered below. Secondly, although multiple scattering will introduce a time dilution effect to the light curve (by delaying the photons and causing them to be detected at later times), we expect it to be small.  The ratio of the contributions from the multiply-scattered component to the singly-scattered one is $\sim \tau \exp{(-\tau)}$; for large optical depths ($\tau \gg 1$), extinction dominates over time dilution, making the multiply-scattered component negligible.

To understand how to deal with initially opaque clouds, we must examine in detail the process of dust sublimation and the relative
time scales.  The former (due to heating) is driven by the burst
flux. Therefore it takes place during the GRB prompt phase, which
lasts $\sim$ 10 s.  The slower IFE process, which is
driven by the burst fluence, lasts for several hours up to a maximum
of a few days.  The dust destruction process takes place
within a thin shell of thickness $\delta R\lesssim10^{15}$ cm, a tiny fraction
of the cloud radius.  Therefore, to a very good approximation, the dust
destruction front propagates at the speed of light. Consider now a
photon that is re-directed towards some
line of sight. Due to the time delay in its propagation, it will be lagging behind the front. As a
consequence, it will propagate only through the post-burst dust population.  To compute the number of photons, a time-dependent dust
treatment is required, but only the final configuration of the cloud is
neccessary to evaluate the extinction of the thermally re-emitted and scattered light.

\begin{figure}
\begin{center}
\includegraphics[width=6in]{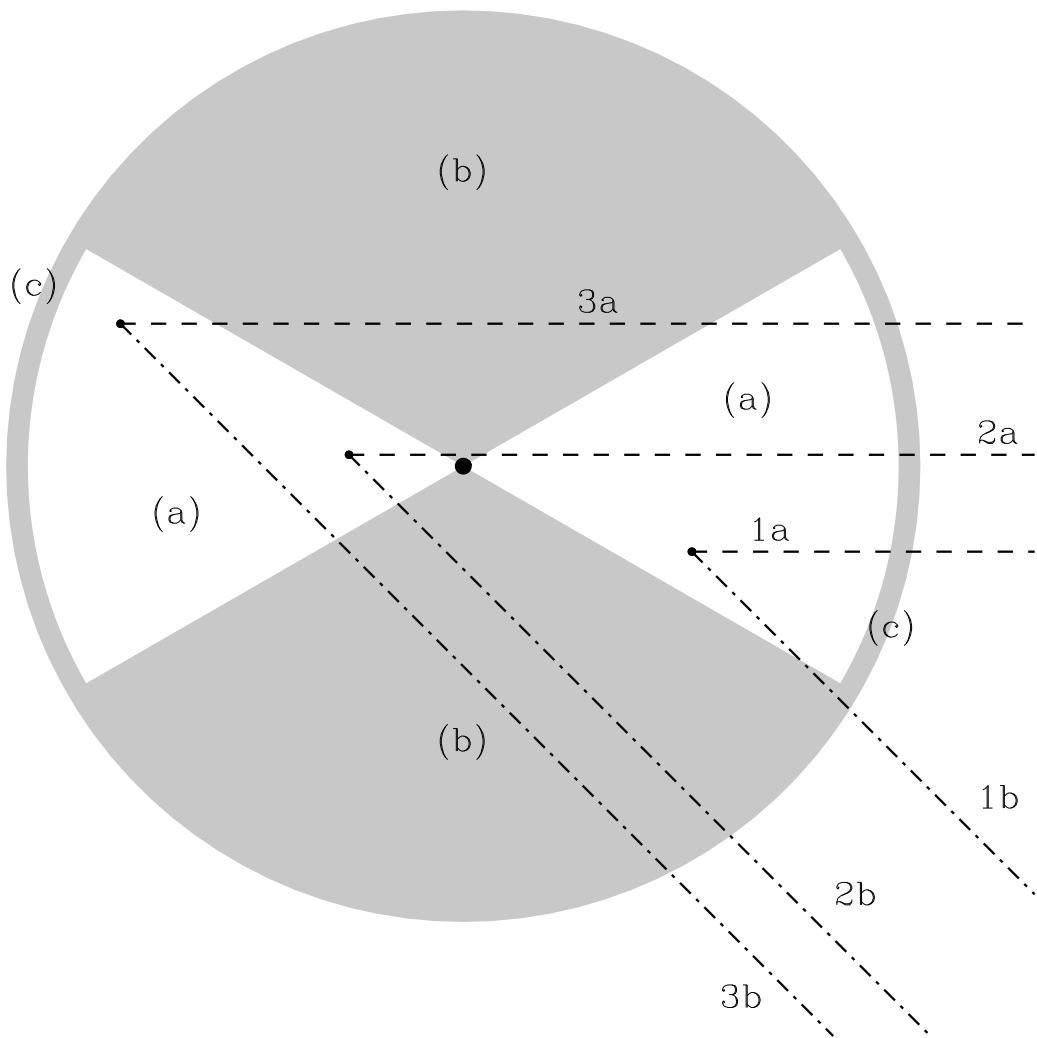}
\end{center}
\caption{Cartoon showing the final configuration of a cloud
illuminated by a beamed GRB --- three regions can be identified in the cloud. Region (a)
has all of the dust completely destroyed by the burst, yielding no residual
opacity. Region (b) is outside the beaming cone and holds the unmodified dust
populations.  Region (c) is a layer of dust left on the
outskirt of the cloud; it is not a general outcome of the simulations as region (a) extends all the way to the edge for more compact clouds. Dashed lines show several paths of re-directed photons from the illuminated dust towards an on-axis observer.  The various
paths range from almost clear (1a) to largely obscured ones (3a). Dot-dashed lines show the paths to an observer located outside the cone --- for such paths, relatively more extinction is generally experienced (cfr. 123a with 123b).
\label{fig:dustcone}}
\end{figure}

This allows us to develop a very simple recipe to deal with isotropic
bursts.  Regardless of the initial opacity in the band considered, we can
compute the echo light curve ignoring the opacity and eventually
correct {\it a posteriori} with the residual extinction (if any).  A
more complex treatment is necessary for beamed GRBs. As shown in
Fig. \ref{fig:dustcone}, a beamed GRB cleans out only a biconical region
of the cloud and, for any initially-opaque cloud, a proper treatment
of the opacity has to be included. A simple one is not possible
in this case since, as shown in the figure, different amounts of dust
lie along different lines of sight. The treatment becomes even more
complicated for off-axis observers, due to the loss of the cylindrical
symmetry. For this reason we restrict our treatment of beamed GRB
echoes to initially optically-thin clouds.

\section{RESULTS}
\label{sect:results}

\subsection{LIGHT CURVES \& SPECTRA}
\label{subsect:lc}

Dust echoes have a duration $t_{\rm{echo}} \sim 2R/c$ and the energy
involved is $E_{\rm{echo}} \sim \zeta E$.  The ``echo efficiency
factor'', $\zeta$, depends on the total amount of dust present in the
scattering and thermally re-emitting region, the distance of the dust
from the burst, the dust composition and the GRB spectrum.  The
dependence of $\zeta$ on the GRB and cloud parameters is complex.  A
large value of $\zeta$ can be achieved if the dust grains are sufficiently far away to survive fast sublimation by the burst photons, yet close
enough to be heated to high temperatures and provide a sizable optical
depth.  Dust grains in the path of the optical-UV flash are expected
to be sublimated out to a distance $\sim 1$ pc (WD00).  Using the PL02
dust code described in \S\ref{subsect:dustcode}, we follow the
evolution of dust grains, with sizes $0.3$ nm $\lesssim a \lesssim
0.3$ $\mu$m, for various values of the density and radius of the
absorbing cloud.

\begin{figure}
\begin{center}
\includegraphics[width=6in]{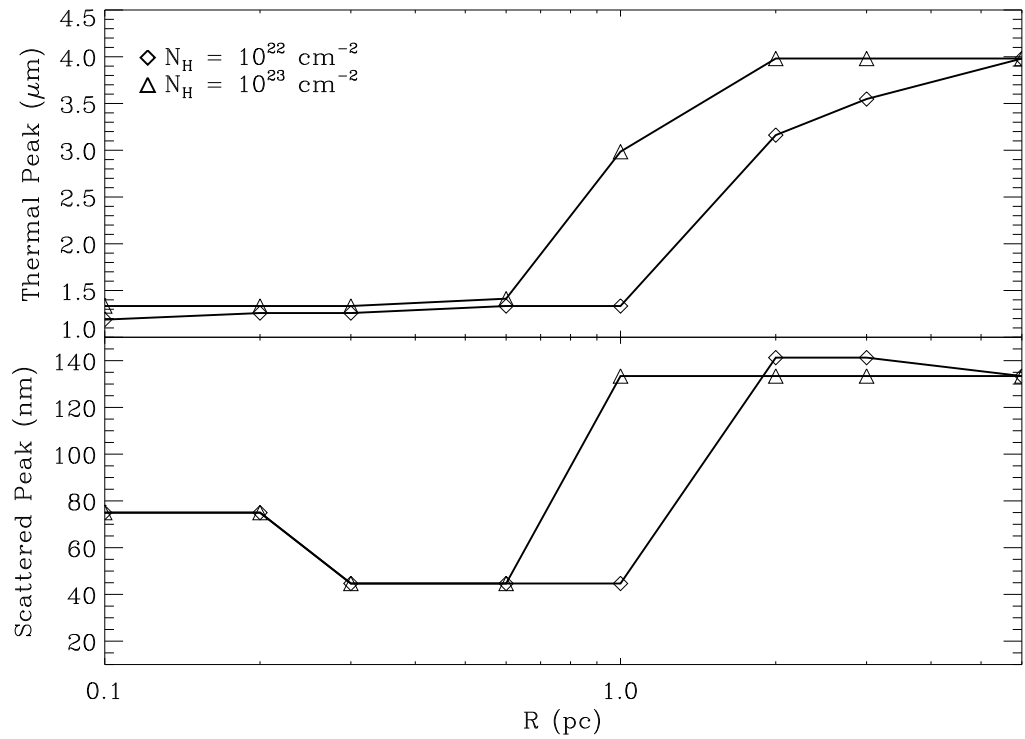}
\end{center}
\caption{Thermal (top) and scattered (bottom) peak wavelengths as a function of the cloud radius $R$ for hydrogen column densities $N_{\rm{H}} = 10^{22}$ and $10^{23}$ cm$^{-2}$.  Note that these are {\it intrinsic} wavelengths for which no extinction correction has been applied to the spectra.}
\label{fig:clouds}
\end{figure}

Figure \ref{fig:clouds} shows how the wavelength of the peak emission,
for the thermal and scattered components, varies as a
function of the cloud parameters. The general trend is for it to increase with cloud size (Fig. \ref{fig:clouds}); since the temperature of the dust declines
with distance from the radiation source, larger clouds have higher
fractions of cooler dust. On the other hand, for a fixed radius, the
peak of the emission is density-independent as long as the cloud is
optically thin (for $N_{\rm
H}=10^{21}$ and $10^{22}$ cm$^{-3}$), but it shifts to longer wavelengths when the cloud
becomes optically thick ($N_{\rm H}=10^{23}$ cm$^{-3}$). This
is due to the fact that if the cloud is optically thick, less radiation
propagates to the outer layers of dust, and therefore the grains
attain a lower temperature.

\begin{figure}
\begin{center}
\includegraphics[width=6in]{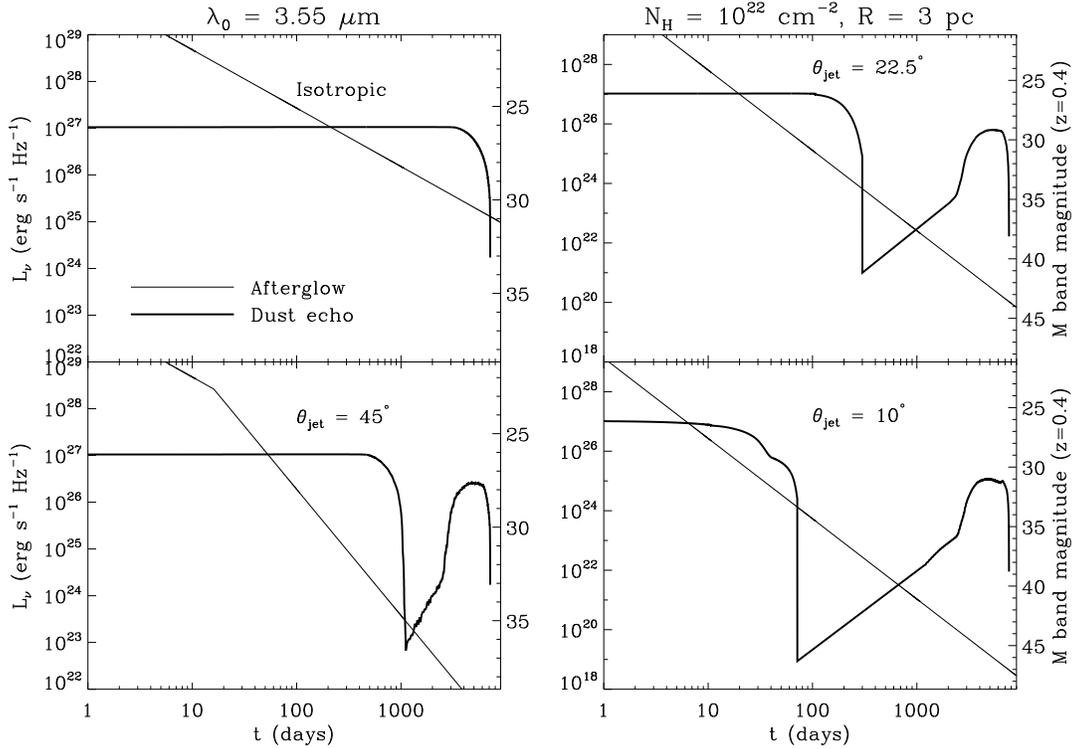}
\end{center}
\caption{Dust echo light curves for on-axis GRBs.  Isotropic and beamed ($\theta_{\rm{jet}} = 10^\circ, 22.5^\circ$ and $45^\circ$) emission are both considered.  The corresponding afterglows are plotted for comparison and $2R/c \sim$ 2400 days.}
\label{fig:onaxis}
\end{figure}

For the remainder of the paper, unless otherwise noted, we will
concentrate on the (optically thin) case of a cloud with radius
$R=3$ pc and $N_{\rm{H}}=10^{22}$ cm$^{-3}$. 
Figure \ref{fig:onaxis} shows echo light curves for the source/rest
wavelength $\lambda_0$ = 3.55 $\mu$m --- we will justify our choice
shortly --- for various values of the jet opening half-angle; the line
of sight is coincident with the axes of symmetric, bipolar jets.  In
all of the cases, the GRB isotropic equivalent energy (in photons) is
$E = 10^{53}$ erg.  For beamed emission, a trough appears in the light
curve at $t\sim R(1-\cos\theta_{\rm{jet}})/c$.  As $\theta_{\rm{jet}}$
decreases, the width of the trough increases.  The figures also show
the afterglow light curves for comparison.  For this particular
configuration, dust echo bumps are visible with a brightness well
above the afterglow, several weeks to months after the GRB.

\begin{figure}
\begin{center}
\includegraphics[width=6in]{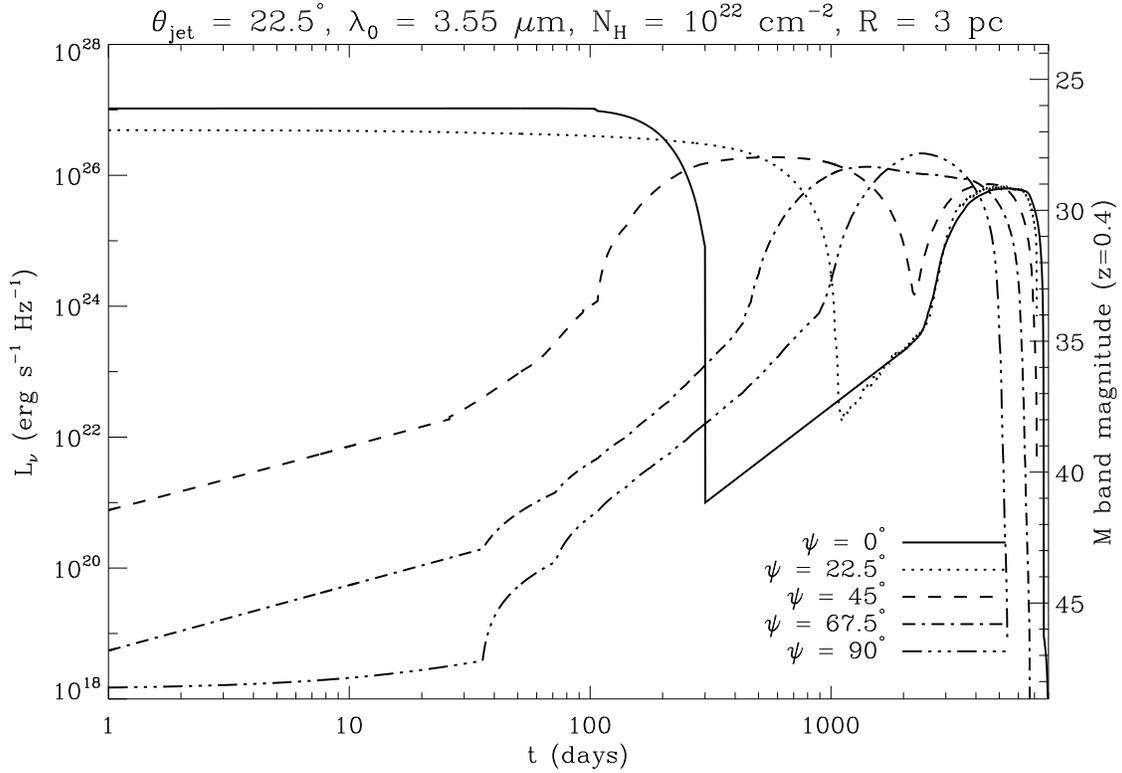}
\includegraphics[width=6in]{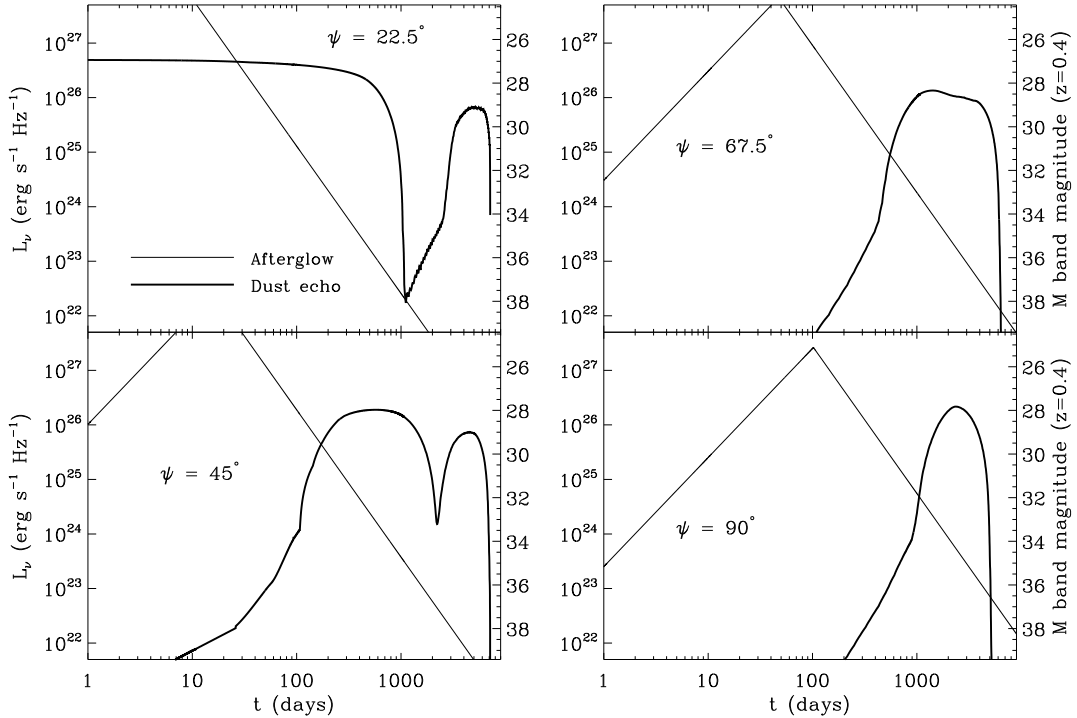}
\end{center}
\caption{Top: Dust echo light curves for jet opening half-angle $\theta_{\rm{jet}} = 22.5^\circ$ and various values of the viewing angle, $\psi$.  Bottom: Light curves with different $\psi$ plotted against the corresponding afterglows.}
\label{fig:viewangles}
\end{figure}

For $\lambda_0 = 3.55$ $\mu$m and $\theta_{\rm{jet}} = 22.5^\circ$, we
compute light curves for different values of the viewing angle
(Fig. \ref{fig:viewangles}).  Even though this value of
$\theta_{\rm{jet}}$ is hardly typical of most GRBs, we adopt it as an
illustration and as a matter of computational expediency.  The main
effect of an increasing $\psi$ is to make the trough shallower, while
lowering the light curve for $t \lesssim R/c$.  Eventually, for $\psi
= 90^\circ$, a bump emerges in place of the trough.  These behaviors
can be understood in the following manner: at $\psi=0^\circ$, the echo
surface at $\sim R/c$ samples the least {\it illuminated} dust; when this configuration
is rotated by $90^\circ$, it now instead samples the most dust.  Note that for Figs. \ref{fig:onaxis} and \ref{fig:viewangles}, we apply a smoothing and interpolation procedure in places where numerical noise is clearly present, taking care not to alter the qualitative behavior of the light curves.

\begin{figure}
\begin{center}
\includegraphics[width=6in]{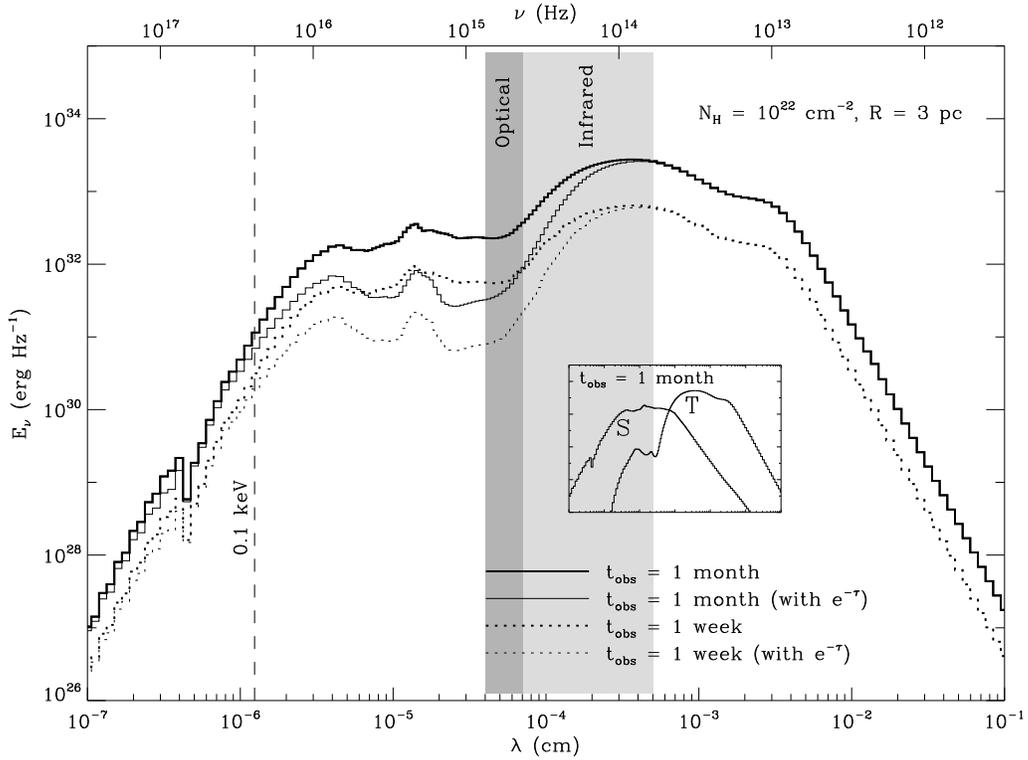}
\end{center}
\caption{Dust echo spectrum integrated from $t=0$ to $t_{\rm{obs}}$.  As an illustration, the insert shows the separated scattered (denoted by ``S'') and thermal (denoted by ``T'') components for $t_{\rm{obs}} =$ 1 month.  Also shown are the spectra with the $\exp{(-\tau)}$ extinction correction applied.}
\label{fig:spectrum}
\end{figure}

For a GRB observed from its onset until a time $t_{\rm{obs}}$, one can
measure the echo energy (per unit frequency),
\begin{equation}
E_\nu = \int^{t_{\rm{obs}}}_0 L_\nu ~dt.
\end{equation}
We compute light curves for the range of rest frame wavelengths 1 nm
$\leq \lambda_0 \leq$ 1 mm, for an isotropic fireball.  Integrating these light curves till
$t=t_{\rm{obs}}$ yields the spectrum, which we obtain for two values of
$t_{\rm{obs}}$ (Fig. \ref{fig:spectrum}).  In the infrared, the echo is dominated by thermal
re-emission from graphite grains.  Since re-emission is isotropic, the
spectrum hardly changes with time.  In the soft X-ray, dust grains
preferentially scatter light forwards, causing the spectrum to be
highly variable, being stronger at early and weaker at late times.  By
separating the dust echoes into thermal and scattered components
(Fig. \ref{fig:spectrum}), we see that they peak at about
$\lambda_{\rm{peak}}$ = 3.55 $\mu$m (infrared) and 8.80 eV (extreme
ultraviolet/soft X-ray), respectively.  The former value justifies our
earlier choice of $\lambda_0 = \lambda_{\rm{peak}}$ in
Figs. \ref{fig:onaxis} and \ref{fig:viewangles}.  Various studies have
yielded different predictions for $\lambda_{\rm{peak}}$: 2 $\mu$m
(EB00), $K$ band (R01) and 8 $\mu$m (Venemans \& Blain 2001).  The
fact that these values differ is hardly surprising, as
$\lambda_{\rm{peak}}$ depends sensitively on the adopted dust
composition, the range of grain sizes and temperatures, as well as the cloud configuration.  It
is worthwhile to note that these calculations are for essentially
static grain populations.  By allowing the dust distribution to
evolve, we include contributions from hotter grains which survive only
for a very short time in the interstellar medium
(Fig. \ref{fig:inputs}).  We also apply an extinction correction, $\exp{(-\tau)}$, due to the presence of a thin, post-burst dust layer, to our spectra; it is clear that this correction is negligible at $\lambda_0 = \lambda_{\rm{peak}}$.

A striking observational feature of dust echoes from GRBs with
symmetric, bipolar jets is the presence of two bumps, each
corresponding to emission from the region illuminated by one of the
two jets.  This is an important difference from SN bumps that are
expected to contribute only a single rebrightening.  Furthermore, the
temporal separation between the two peaks allows one to directly infer
the opening half-angle, since the first bump appears at $t_1\sim
R(1-\cos\theta_{\rm{jet}})/c$ while the second emerges at $t_2\sim
2R/c$.  For a narrow jet, we have
\begin{equation}
\theta_{\rm{jet}} \sim 2\sqrt{\frac{t_1}{t_2}}.
\label{eq:thetajet}
\end{equation}
If the GRB does not sit in the center of the cloud, equation
(\ref{eq:thetajet}) becomes more complicated and we then need to use
the relative brightnesses of the peaks to constrain the burst
location.

\subsection{OBSERVATIONAL PROSPECTS}
\label{subsect:observations}

For a region of size $\sim$ 3 pc and density
$n_{\rm{H}} \sim  10^3$ cm$^{-3}$, the luminosity of the dust echoes is $\sim 10^{27}$ erg s$^{-1}$ Hz$^{-1}$ at a
wavelength of 3.55 $\mu$m, allowing them to sit above the
afterglow light curve for time scales of about a month to over a year.

\begin{figure}
\begin{center}
\includegraphics[width=6in]{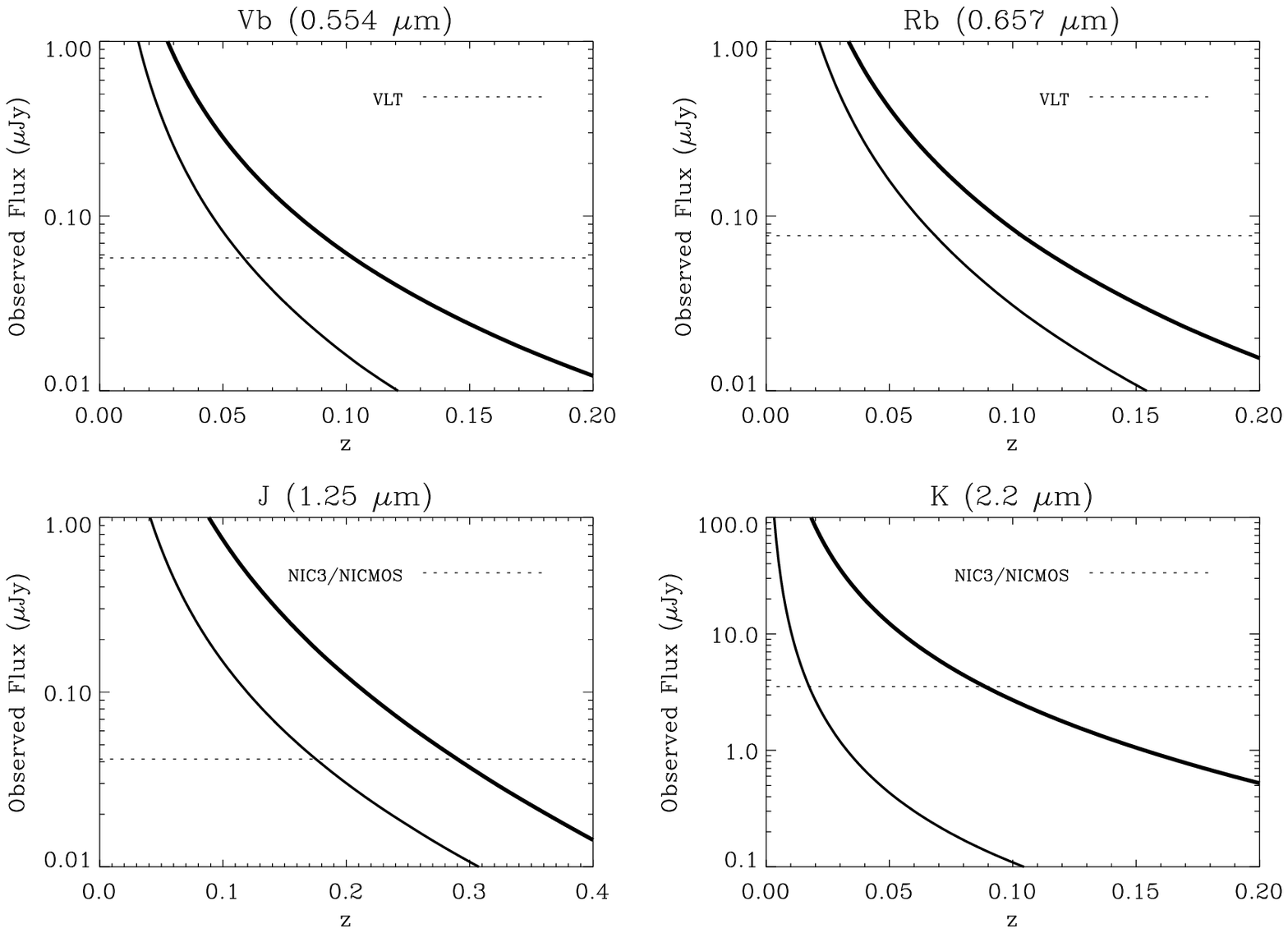}
\includegraphics[width=6in]{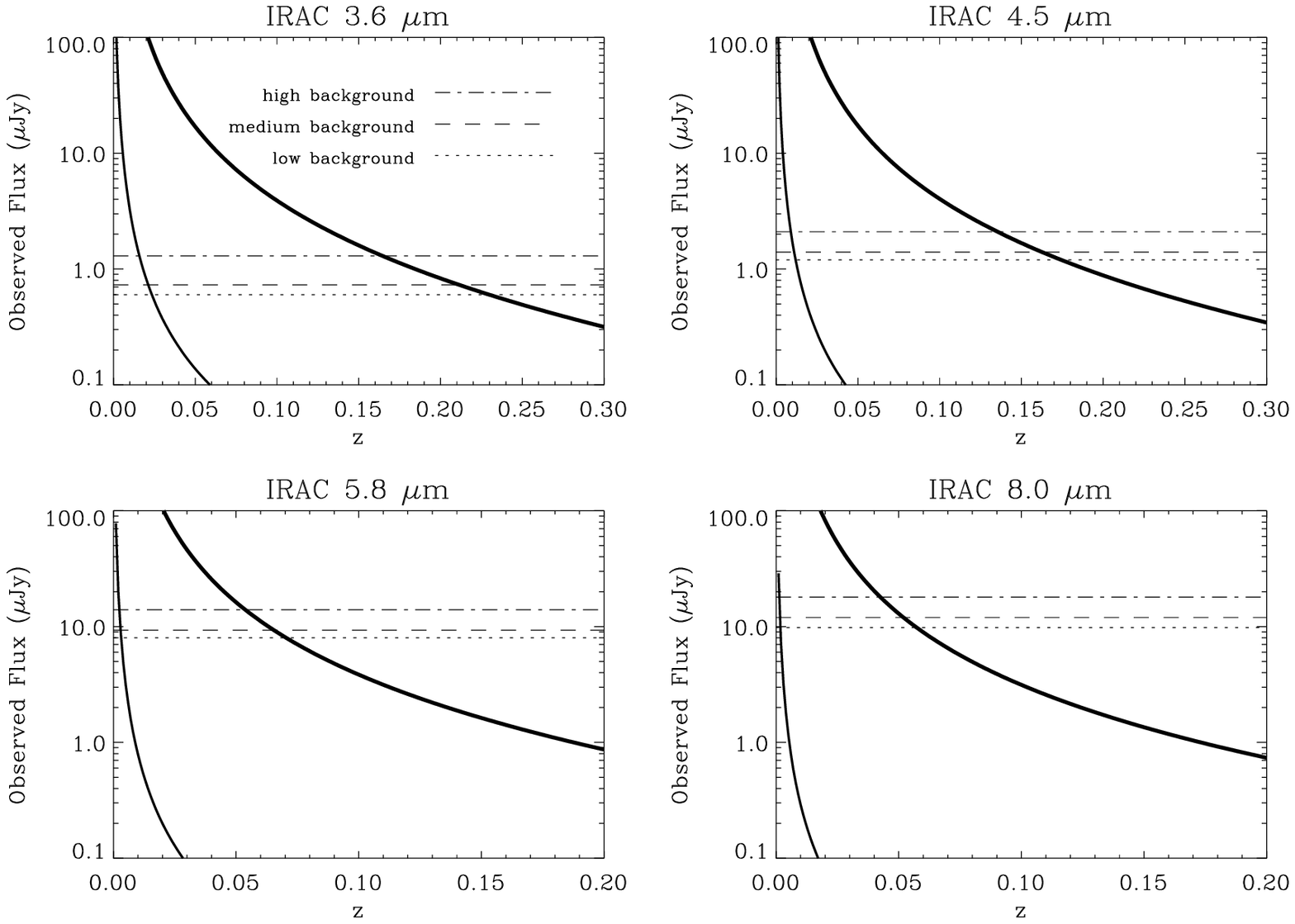}
\end{center}
\caption{Time-averaged, observed echo flux, from isotropic fireballs, as a function of redshift, $z$, with the extinction correction applied.  Shown are two $N_{\rm{H}} = 10^{22}$ cm$^{-2}$ cloud models for comparison, with the thin and thick lines representing $R=1$ and 3 pc, respectively.  We display the various minimum flux thresholds for:  both the $Vb$ and $Rb$ bands of the {\it VLT} (S/N $\sim$ 5, exposure time $t_{\rm{exp}} \sim 1$ hr); NIC3/NICMOS onboard {\it Hubble} (S/N = 5, $t_{\rm{exp}} = 3600$ s); and the four channels of IRAC onboard {\it Spitzer} (for 100 s frame times).}
\label{fig:obs}
\end{figure}

In Figs. \ref{fig:onaxis} and \ref{fig:viewangles}, we display
additional axes for the $M$ magnitude of the echoes.  At $\lambda_0$ =
3.55 $\mu$m, we need the source to be located at $z \approx 0.4$ in order to
detect its light in the $M$ band (5 $\mu$m).  For on-axis GRB jets,
the echo bumps have a magnitude of $M \sim 26-27$, straining the
detection limits of ground-based telescopes, but well within the reach
of IRAC onboard the {\it Spitzer Space Telescope}.  Specifically, for a cloud of
$R=3$ pc and $N_{\rm{H}}=10^{22}$ cm$^{-2}$, infrared echoes should be detectable with IRAC at 3.6 and 4.5 $\mu$m up to $z\sim 0.2$ (Fig. \ref{fig:obs}) and with NICMOS (in the $J$ band) onboard the {\it Hubble Space Telescope} up to $z \sim 0.3$.

In the optical, the echoes are less bright, with expected magnitudes
of $Rb \sim 30$ and $Vb \sim 30$.  Figure \ref{fig:obs} shows minimum flux thresholds for the {\it Very Large Telescope (VLT)} and demonstrates that optical echoes can be seen up to $z \sim 0.1$.  Finally, in the X-ray, the echo flux is
dominated by scattering --- in the 0.1 to 1 keV band, we have $F_X \sim
10^{-16}$ erg cm$^{-2}$ s$^{-1}$ at $z \sim 0.1$.  Such a low flux is
barely detectable with a long {\it Chandra} exposure, making the X-ray
band an unattractive prospect in the search for dust echoes from the
circumburst medium.

Extrapolating these results to different dust geometries is not
straightforward. Naively, the dust luminosity scales as
\begin{equation}
L_{\rm{echo}} \sim \zeta \frac{cE}{2R},
\label{eq:wing}
\end{equation}
where $\zeta$ hides the complexity of the time-dependent dust
distribution.  Equation (\ref{eq:wing}) gives an optimistic scaling
and should be used with caution. On one hand, $\zeta$ will be low for
a very large cloud, since the grains will not reach very high
temperatures.  However, a very small region will also have a low
$\zeta$ due to the fact that all of the dust is sublimated in the very
first few fractions of a second by the burst radiation.  Increasing
the amount of dust increases $\zeta$, but at the expense of causing
residual absorption by dust in the cloud, delaying the echoes and
thereby lowering their luminosity.  
\section{DISCUSSION}
\label{sect:discussion}

We have computed dust echoes due to scattering and thermal re-emission
of dust by GRB sources.  Our model allows for time-dependent dust
destruction by both isotropic and beamed radiation from the GRB, as
well as for any inclination of the jet axis with respect to the line
of sight to the observer.  Our source model assumes that the outer
edges of the jets are sharp. Under these conditions and the assumption
of symmetric, bipolar jets, a pair of bumps is expected in the light
curve, an intrinsically different behavior from the possible presence of a single SN bump.  This
feature should qualitatively remain even if the jet is structured
(Rossi, Lazzati \& Rees 2002; Zhang \& M\'{e}sz\'{a}ros 2002).  The
energy input in this case is a function of the angle from the jet axis
--- namely, $\epsilon \propto \Theta^{-k}$, where we typically have
$1.5 < k \lesssim 2$.  [However, we note that the numerical
simulations of Morsony, Lazzati \& Begelman (2006) allow for $k \gg
1$.]  Due to the energy dependence with angle from the jet axis, dust
is destroyed out to larger radii along the jet axis than in its wings.
The range of grain temperatures present is therefore expected to vary
with azimuthal angle.  The light travel paths of the photons differ,
due to different geometries created by the destruction fronts. Thus,
one expects varying observational features depending on the structure
of the GRB jet.

Qualitatively, one can think of a structured jet as a superposition of
a number of {\it non-structured} jets with different
$\theta_{\rm{jet}}$.  The component with the smallest opening
half-angle, $\theta_{\rm{core}}$, forms the core of the structured
jet, produces the brightest echo and displays a trough in its light
curve at the earliest time.  The net effect from summing all of the
components is to form an echo light curve with a gentle decline at
$t_{\rm{core}} \sim R(1 - \cos\theta_{\rm{core}})/c$, rather than a
sharp transition to a trough.  As a consequence, the first echo bump
will likely be undetectable, since it lies at very early times and is
outshone by the afterglow.

Observationally, we conclude that even echoes from moderate redshift
events are hard to detect with present ground-based telescopes.
Infrared observations with {\it Spitzer} and {\it Hubble} --- and in
the future with the {\it James Webb Space Telescope} --- will
detect echo components in GRBs out to $z \sim 0.3$.  An intriguing
prospect is the detection of orphan dust echoes, which are those from
a misaligned jet.  They are almost as bright as echoes from an on-axis
GRB, but are not contaminated by the burst and afterglow radiation.  Detecting such echoes in the optical light curves of supernovae may serve as the observational signature of off-axis, ``phantom'' GRBs (and their accompanying ``orphan afterglows''), which are otherwise exceedingly difficult to detect.

\acknowledgments 

We wish to thank: Peter Ruprecht and James McKown for steadfast and
unwaivering technical support at JILA; Brian Morsony for illuminating
discussions concerning structured jets; Enrico Ramirez-Ruiz for useful conversations regarding orphan afterglows; Richard
McCray and Svetozar Zhekov for helpful comments and suggestions, following their careful
reading of the manuscript. This work was partially supported by NASA
Grants NAG5-12279 (K.H.) and NAS5-26555 (K.H.), and NSF Grant AST 0507571 (K.H., D.L. and R.P.).


\end{document}